\documentclass[twocolumn]{aastex62}

\usepackage{color}
\usepackage{graphicx}
\usepackage{lineno}

\shorttitle{Observational approach to detect ${\rm O_2}$}
\shortauthors{Lopez-Morales et al.}

\begin{document}

\title{Optimizing Ground-based Observations of ${\rm O_2}$ in Earth Analogs}

\correspondingauthor{Mercedes Lopez-Morales}
\email{mlopez-morales@cfa.harvard.edu}

\author[0000-0003-3204-8183]{Mercedes Lopez-Morales}
\affil{Center for Astrophysics ${\rm \mid}$ Harvard {\rm \&} Smithsonian, 60 Garden Street, Cambridge, MA 02138, USA}
\author[0000-0001-6760-3074]{Sagi Ben-Ami}
\affil{Center for Astrophysics ${\rm \mid}$ Harvard {\rm \&} Smithsonian, 60 Garden Street, Cambridge, MA 02138, USA}
\author[0000-0002-8090-6480]{Gonzalo Gonzalez-Abad}
\affil{Center for Astrophysics ${\rm \mid}$ Harvard {\rm \&} Smithsonian, 60 Garden Street, Cambridge, MA 02138, USA}
\author[0000-0003-1361-985X]{Juliana Garcia-Mejia} \altaffiliation{National Science Foundation Graduate Research Fellow}
\altaffiliation{ 
Ford Foundation Fellow}
\affil{Center for Astrophysics ${\rm \mid}$ Harvard {\rm \&} Smithsonian, 60 Garden Street, Cambridge, MA 02138, USA}
\author[0000-0001-6320-7410]{Jeremy Dietrich}
\affil{Center for Astrophysics ${\rm \mid}$ Harvard {\rm \&} Smithsonian, 60 Garden Street, Cambridge, MA 02138, USA}
\affil{Department of Astronomy and Steward Observatory, University of Arizona, 933 N Cherry Ave, Tucson, AZ 85719, USA}
\author[0000-0002-0255-2525]{Andrew Szentgyorgyi}
\affil{Center for Astrophysics ${\rm \mid}$ Harvard {\rm \&} Smithsonian, 60 Garden Street, Cambridge, MA 02138, USA}



\begin{abstract}

We present the result of calculations to optimize the search for molecular oxygen, ${\rm O_2}$ in Earth analogs transiting around nearby, low-mass stars using ground-based, high-resolution, Doppler shift techniques. We investigate a series of parameters, namely spectral resolution, wavelength coverage of the observations, and sky coordinates and systemic velocity of the exoplanetary systems, to find the values that optimize detectability of ${\rm O_2}$. We find that increasing the spectral resolution of observations to R $\sim$ 300,000 - 400,000 from the typical R $\sim$ 100,000, more than doubles the average depth of ${\rm O_2}$ lines in planets with atmospheres similar to Earth's. Resolutions higher than $\sim$500,000 do not produce significant gains in the depths of the ${\rm O_2}$ lines. We confirm that observations in the ${\rm O_2}$ A-band are the most efficient except for M9V host stars, for which observations in the ${\rm O_2}$  NIR-band are more efficient. Combining observations in the ${\rm O_2}$ A, B, and NIR -bands can reduce the number of transits needed to produce a detection of ${\rm O_2}$ by about 1/3 in the case of white noise limited observations. However, that advantage disappears in the presence of typical levels of red noise. Therefore, combining observations in more than one band produces no significant gains versus observing only in the A-band, unless red-noise can be significantly reduced. Blending between the exoplanet's ${\rm O_2}$ lines and telluric ${\rm O_2}$ lines is a known problem. We find that problem can be alleviated by increasing the resolution of the observations, and by giving preference to targets near the ecliptic.

\end{abstract}

\keywords{methods: observational ---  planets and satellites: atmospheres, composition, detection, terrestrial planets --- techniques: radial velocities}


\section{Introduction} \label{sec:intro}

The appearance of molecular oxygen, ${\rm O_2}$,  in large amounts in Earth's atmosphere has been linked for decades to the onset of oxygenic photosynthesis by early organisms \citep[see overview in][]{Catling2017}. The ${\it Galileo}$ probe observations by \cite{Sagan1993}, where they observed strong absorption features of telluric ${\rm O_2}$ and methane, which departed from local thermodynamical equilibrium, drew further attention to those gases as a strong indicator of life in our planet.


The idea of searching for  ${\rm O_2}$ in terrestrial planets around nearby stars has gained additional momentum in the past few years. Given the recent discoveries of Proxima b \citep{AngladaEscude2016}, the Trappist-1 planets \citep{Gillon2017}, LHS 1140 b  \citep{Dittmann2017}, Ross 128 b \citep{Bonfils2018}, and the expected outcome of the TESS mission \citep{Ricker2016} -- predicted to find about 10 planets with radii smaller than two Earth radii within or near the habitable zone of nearby stars \citep{Barclay2018} -- finding the best way to detect  ${\rm O_2}$ in the atmosphere of those planets is becoming a  pressing task. There is also known abiotic pathways to produce ${\rm O_2}$ on Earth-like planets \citep[e.g][]{Domagal2014,Wordsworth2014,Luger2015}. Those potential scenarios enhance the importance of detecting ${\rm O_2}$ in Earth analogs, so we can start to understand how diverse the atmospheric characteristics of Earth-analogs truly are.

\cite{Snellen2013} first proposed to use high-resolution (R = 100,000), Doppler shift observations centered on the ${\rm O_2}$ A-band at 0.76 microns, and cross-correlation techniques to detect  ${\rm O_2}$ in the atmosphere of transiting Earth twins. They concluded this could be achieved with instrumentation on the upcoming generation of Extremely Large Telescopes (ELTs). Particularly, they concluded that the 39-meter European Extremely Large Telescope (ELT) would be able to detect this ${\rm O_2}$ signal in the transmission spectrum of an Earth-like planet transiting an M5V star, 12 parsecs away, by combining the observations of 30 transits. They also pointed out that for planets around M7V stars or later, it would be easier to detect ${\rm O_2}$ by observing around the ${\rm O_2}$ band centered at 1.27 microns, where late M dwarfs are brightest.

\cite{Rodler2014} [hereafter RLM14] refined the \cite{Snellen2013} calculations by producing simulated observations which included the effects of Poisson noise and correlated (red) noise, typically introduced 
by instrumental and atmospheric effects, and by also accounting for atmospheric refraction
effects in the exoplanet's transmission spectrum \citep{GarciaMunoz2012}. In addition, RLM14 explored the expected performance of high resolution spectrographs planned for the
three upcoming ELTs (GMT, TMT, and E-ELT), including a range of spectral resolutions determined by slit-width, and a range of stellar spectral types and their velocities with respect to Earth. They concluded that, although still feasible, the number of transits needed to detect ${\rm O_2}$ in an Earth-twin around a nearby
M dwarf was two times larger than predicted by \cite{Snellen2013} after accounting for red noise. They also found that the ${\rm O_2}$ A-band at 0.76  ${\rm \mu m}$ is better than the ${\rm O_2}$ NIR-band at 1.27 ${\rm \mu m}$ to detect the exoplanet's signal for all M dwarf spectral types, except M9V. This is because instrumental and atmospheric noise levels are typically larger in the near-IR. RLM14 also found that blending effects between telluric ${\rm O_2}$ lines and the exoplanet's ${\rm O_2}$ lines limits the number of targets well suited for this type of observations. This problem appeared most significant in the case of systems with relative velocities of less than $\pm$ 15 km/s and around $\pm$ 50 km/s with respect to Earth, where 50$\%$ or more of the exoplanet ${\rm O_2}$ lines appeared blended.

More recently \cite{Serindag2019} generated more realistic simulations of noise effects by injecting a simulated ${\rm O_2}$ signal on to real time series observations of Proxima Centauri with the UVES instrument on the VLT. By rescaling the flux of Proxima Centauri collected by the VLT to the flux of a hypothetical M5V star 7 parsecs away hosting an Earth twin, they find that ${\rm O_2}$ can be detected in 20 -- 50 transits. That result is consistent with the predictions by \cite{Snellen2013} and RML14. 

RLM14 looked into what would be the optimal spectral resolution of a standard high-resolution slit spectrograph to detect ${\rm O_2}$ and found it to be R = 80,000 once slit losses and the smearing of the signal as resolution decreases are taken into account. They also suggested that if more star light could be fed into a narrow slit using pre-slit optics, the efficiency of observations at higher resolutions can greatly improve (see Fig.~3 in RLM14). Here we show how the efficiency of the observations can be improved even further by increasing the spectral resolution of the signal to values that optimize the depth of the ${\rm O_2}$ spectral lines, making the signal easier to detect.

In this paper we explore what are the best values for a set of instrumental and observational parameters to optimize the detectability of ${\rm O_2}$ in Earth-like exoplanets. Those parameters are: spectral resolution (Section 2), combination of absorption bands (Section 3), and the most suitable set of targets to minimize/circumvent the line-blending problem between exoplanetary and telluric ${\rm O_2}$ lines (Section 4). Sections 5 and 6 contain the summary and discussion of our results.

\section{Spectral Resolution} \label{sec:resolution}

In Earth's atmosphere ${\rm O_2}$ extends over a large range of atmospheric heights,  from the surface of the planet to altitudes of 85 -- 100 km \citep[see e.g.][]{Misra2014}, therefore the characteristic shape of the ${\rm O_2}$ lines is produced by absorption of stellar photons over a wide range of atmospheric temperatures and pressures. Given our current one-planet sample, we do not know how typical the distribution of ${\rm O_2}$ observed in the Earth atmosphere is, but for the purpose of our simulations, we assume Earth-like exoplanets have a similar vertical distribution of ${\rm O_2}$ as Earth. Each ${\rm O_2}$ spectral line has a finite width and a depth, which are determined by the effect of two broadening mechanisms: pressure broadening in the high-pressure layers of the atmosphere, and Doppler broadening in the lower pressure layers \cite[see e.g.][]{Chance2017}. The combination of those two broadening mechanisms defines a minimum width and a maximum depth for the spectral lines, beyond which, instrumentation with resolving power larger than the intrinsic broadening of the lines will not improve our capacity to detect those ${\rm O_2}$ lines.

We performed a series of simulations to establish at what resolution ${\rm O_2}$ lines appear fully resolved in the transmission spectrum of a planet like Earth. We simulated spectra with resolutions ${\rm R = 10^5 - 10^6}$, in ${\rm 10^5}$ steps using the Reference Forward Model \citep[RFM\footnote{RFM is a GENLN2-based \citep{Edwards1992}, general purpose line-by-line radiative transfer model originally developed at Oxford University, under an ESA contract to provide reference spectral calculations for the Michelson Interferometer for Passive Atmospheric Sounding (MIPAS) instrument \citep{Fischer2008}.};][]{Dudhia2017}, over a wavelength range between 0.57 and 1.35 microns, which includes four ${\rm O_2}$ bands observed in the optical and near-infrared transmission spectrum of Earth's atmosphere: the ${\rm \alpha}$ band at 0.63  ${\rm \mu m}$, the ${\rm B}$ band at 0.69 ${\rm \mu m}$, the ${\rm A}$ band at 0.76 ${\rm \mu m}$,  and the ${\rm NIR}$ band at 1.27 ${\rm \mu m}$. In our simulations the RFM is driven by the HITRAN 2012 spectral database \citep{Rothman2013}. We used the US standard atmosphere (1976)\footnote{https://ccmc.gsfc.nasa.gov/modelweb/atmos/us$\_$standard.html} to simulate temperature, pressure, and molecular oxygen and water vapor profiles from the surface up to 100 km at 2 km intervals (or layers). To produce the total transmission spectrum of the exoplanet atmosphere, ${\rm T_T}$, we integrated  the transmission of each layer, ${\rm T_n}$, as 
\begin{equation}
{\rm T_T} = \frac{\rm \sum~T_n A_n }{\rm A_T} ,
\end{equation} 
\noindent where ${\rm A_n}$ the cross-section area of each layer as seen from Earth, and ${\rm A_T}$ is the total cross-section area of the atmosphere, i.e. ${\rm A_T}$ = ${\rm \sum~A_n}$. In this integration, we accounted for the effect of refraction in the bottom layers of the atmosphere, as described in RLM14. Specifically, we did not include the first three  model layers, between 0 km and 6 km from the exoplanet's surface, which are the layers affected by refraction in planets in the habitable zone of early M dwarfs \citep{GarciaMunoz2012}. We made the same assumption for all host star spectral types, although refraction might have a lesser effect in late-type M-dwarfs \citep[see e.g.][] {Betremieux2014}.

\begin{figure*}[h!t]
\centering
\includegraphics[trim=1.0cm 6cm 0cm 3cm,clip=true,width=\textwidth]{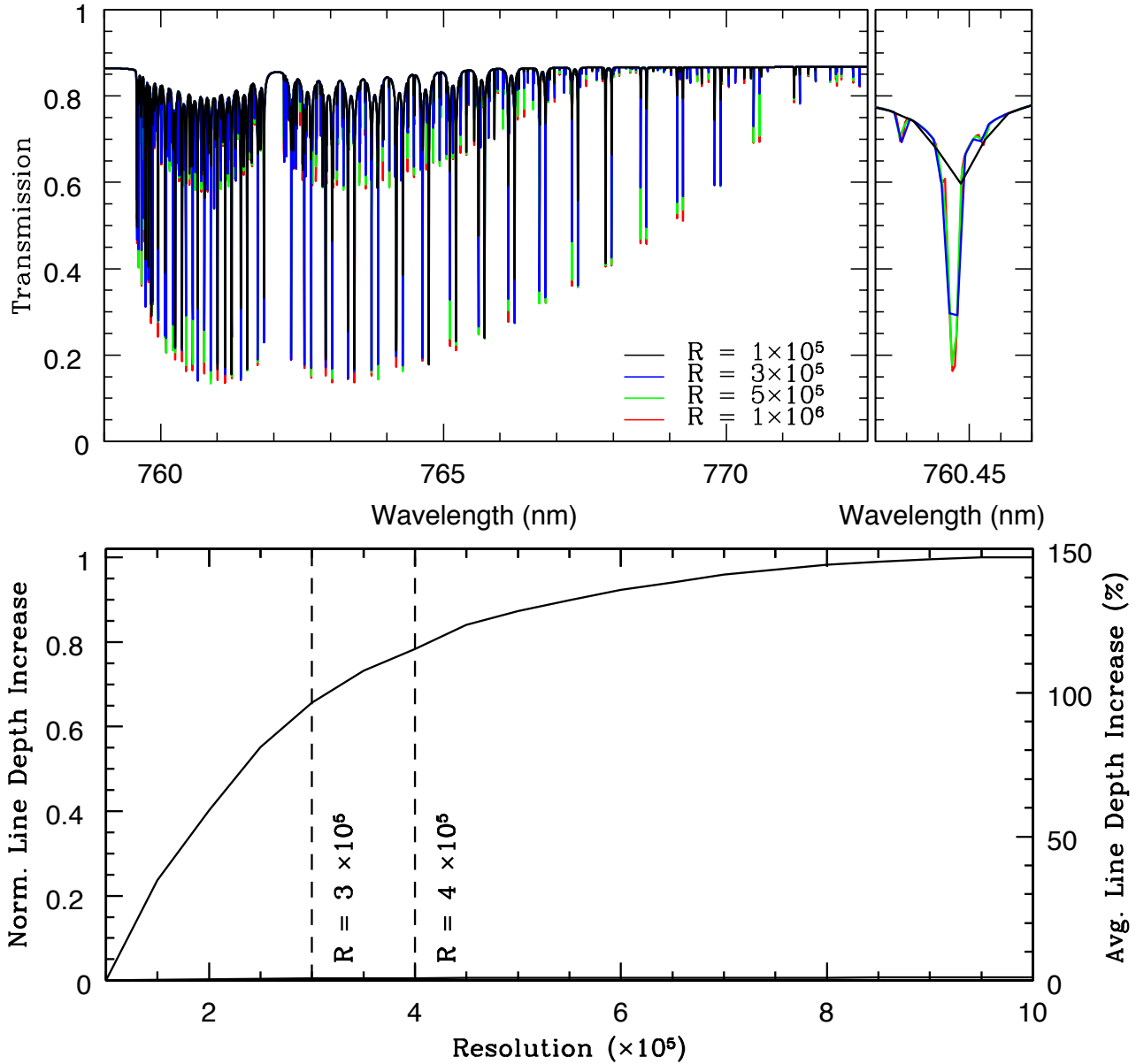}
\caption{{\it Top} -- 0.76 $\mu$m ${\rm O_2}$ band at four different resolutions: R = 100,000 (black), 300,000 (blue), 500,000 (green), and 1,000,000 (red). The right-side panel shows a zoom-in of one of the lines in the band. {\it Bottom} -- Increase in integrated depth of the ${\rm O_2}$ lines as a function of resolution, with respect to R = 100,000 and normalized to the integrated depth at resolution R = 1,000,000.\label{fig:Aband}}
\end{figure*}

\begin{figure*}[h!t]
\centering
\includegraphics[trim=1.0cm 6cm 0cm 3cm,clip=true,width=\textwidth]{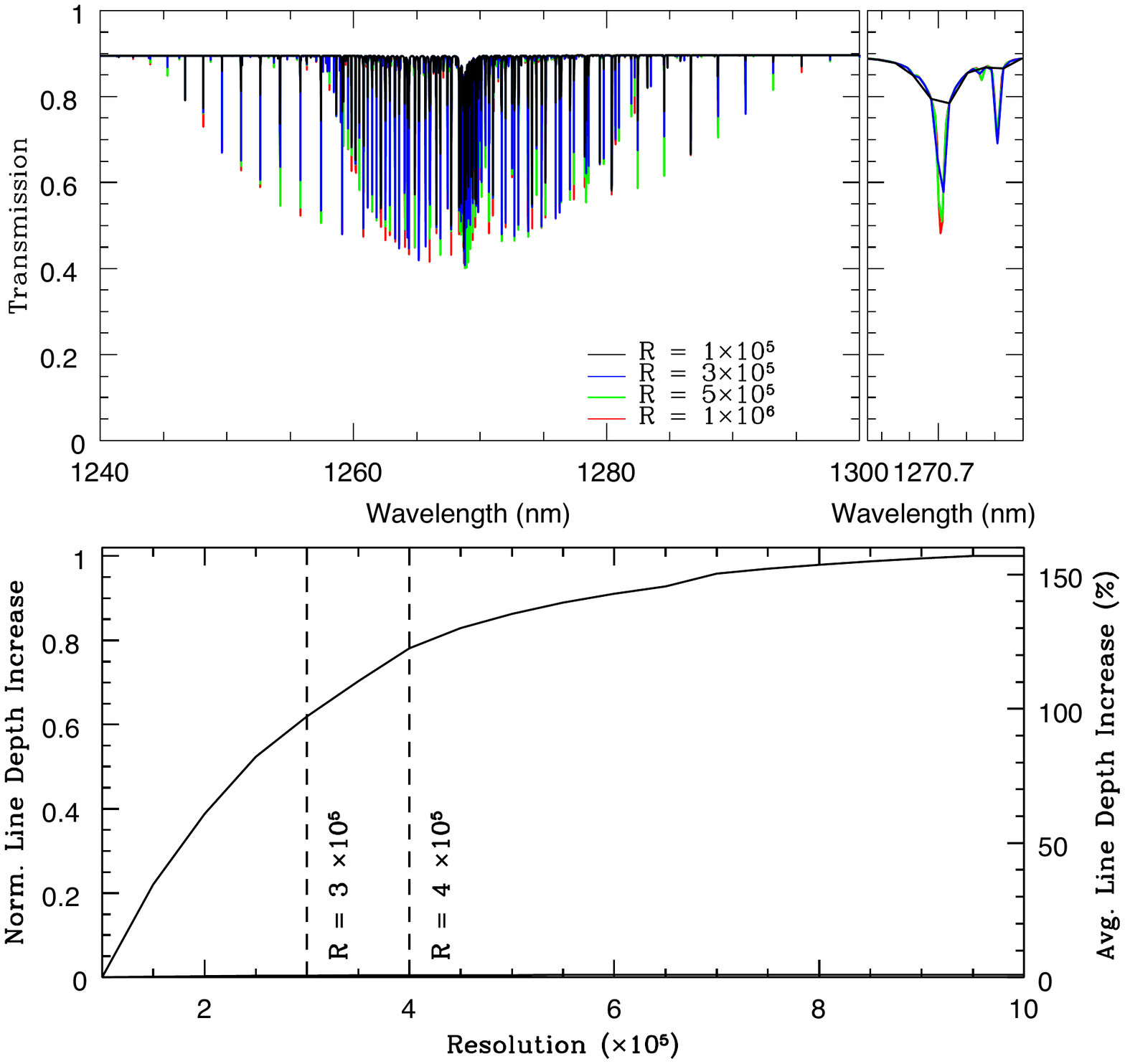}
\caption{Same as Figure~\ref{fig:Aband}, but for the 1.27 $\mu$m ${\rm O_2}$ band.\label{fig:00band}}
\end{figure*}

We show sample results of these simulations in the top panels of Figures 1 and 2, where we compare transmission spectra in the 0.76 ${\rm \mu m}$ ${\rm A}$ band, and the 1.27 ${\rm \mu m}$ ${\rm NIR}$ band at different resolutions. The bottom panels in each figure show how the integrated depth of the ${\rm O_2}$ lines increases as a function of resolution for each of those two bands. The integrated line depth for each resolution is calculated by first identifying all the lines in each spectra with depth larger than 5$\%$ with respect to the local continuum, and then adding their individual depths. Each line depth is calculated as the minimum of a Gaussian fit to the line.


\noindent The vertical axes in the bottom panels of Figures 1 and 2 show the normalized integrated line depths at each resolution, ${\rm I_R}$, using the integrated line depth at ${\rm R = 1\times10^5}$, ${\rm I_{10^5}}$, as zero point, and normalizing by the integrated line depth at ${\rm R = 1\times10^6}$, ${\rm I_{10^6}}$. That is

\begin{equation}
 {\rm Norm.~Integrated~Depth} = \frac{({\rm I_R} - {\rm I_{10^5}})}{({\rm I_{10^6}} - {\rm I_{10^5}})} .
\end{equation}

\noindent Those figures show how we gain between 100${\rm \%}$ and 
150${\rm \%}$ in line depths by increasing the instrumental resolution from ${\rm R = 1\times10^5}$  to ${\rm R = 3\times10^5 - 4\times10^5}$. Given the multiplicative nature of cross correlation functions (CCFs), an increase in the depth of the signal translates into an increase in the strength of the CCFs peaks, making the signal easier to detect. The ${\rm O_2}$ lines appear fully resolved at resolutions larger than about ${\rm R = 8.5\times10^5}$, and increasing instrumental resolution past ${\rm R = 4\times10^5}$ only produces about 20${\rm \%}$ additional gain in line depths.
The same calculations for the ${\rm O_2}$  ${\rm \alpha}$-band at 0.63 ${\rm \mu m}$ and the ${\rm B}$-band at 0.69 ${\rm \mu m}$ produce similar results.

\section{Combination of ${\rm O_2}$ Bands} \label{sec:wavelength}

As shown in the original detections of CO in hot Jupiters by \cite{Snellen2010}, \cite{Brogi2012}, and \cite{Rodler2012}, our ability to detect a particular molecular species in an exoplanet atmosphere using high-resolution spectroscopy is enhanced by the capacity of cross-correlation techniques to simultaneously use information from multiple absorption lines, ${\rm N_l}$. Work done on hot Jupiters so far, and also the simulation studies done on the detectability of ${\rm O_2}$ in Earth analogs \citep{Snellen2013,Rodler2014,Serindag2019}, have focused on individual bands of specific molecules. However, given that the strength of the CCF signal increases as ${\rm \sqrt{N_l}}$ \cite[see eq.~1 in][]{Snellen2015}, it might be advantageous to observe multiple bands simultaneously, or to combine observations in different bands. In this section we explore whether multi-band observations can improve our ability to detect ${\rm O_2}$. Of the four ${\rm O_2}$ bands between 0.6 ${\rm \mu m}$ and 1.3 ${\rm \mu m}$, the A band (0.76 ${\rm \mu m}$) is the deepest, followed by the B band (0.69 ${\rm \mu m}$), and the ${\rm NIR}$ band (1.27 ${\rm \mu m}$) with similar depths. The ${\rm \alpha}$ band (0.63  ${\rm \mu m}$) is significantly weaker than the other three, therefore we do not use that band in the rest of this study. 
Using the ${\rm A}$, ${\rm B}$, and ${\rm NIR}$ bands, we produce as realistic as possible simulations for different combinations of those bands, assuming exoplanet host stars of different spectral types, and typical noise levels. In the simulations, we adopt a spectral resolution  ${\rm R = 3\times10^5}$, based on the results in Section 2, and place the hypothetical planetary system at a fixed distance of 5 parsecs from Earth. 

\subsection{Generation of model spectra}
We generated model spectra following the same methodology described in RLM14 using their eqs. (1) and (2), which we reproduce here. The model spectra ${\rm C}$ are given by

\begin{equation} \label{E3}
C=\Big(a~(1+\epsilon^{-1})^{-1} T \Big) \otimes G~,
\end{equation} 
with
\begin{equation} \label{E4}
a=(1+v_\star c^{-1})~S + \left(1+(v_\star+v_{\rm pl})c^{-1} \right) \epsilon^{-1}~P~,
\end{equation} 

\noindent where ${\rm S}$ is the spectrum of the host star, ${\rm P}$ is the transmission spectrum of the planet, ${\rm T}$ is the telluric spectrum of Earth, and ${\rm G}$ is the instrumental profile of the instrument used to perform the observations. The other parameters in the equations are the speed of light ${\rm c}$, the instantaneous radial velocity of the planet with respect to its host star ${\rm v_{pl}}$ (${\rm \sim 0~km~s^{-1}}$ at primary transit), the relative velocity of the system with respect to Earth ${\rm v_{*}}$, and the ratio of areas between the stellar disk and the atmospheric ring of the planet ${\rm \epsilon}$.  

For the host star models, we focused on M-dwarf stars following previous studies \citep{Kaltenegger2009,Snellen2013,Rodler2014}, which show that the exoplanet atmospheric signal is larger for planets around late type stars. This is because the amplitude of those signals is proportional to the ratio of the areas of the stellar disk and the ring subtended by the atmosphere of the planet. In addition, observing Earth-like planets around M-dwarf stars has the advantage that a large number of transit observations, needed to detect atmospheric features, can be obtained in a relatively short period of time.

We used three stellar emission models with spectral types M1V, M4V, and M9V to explore the effect of relative stellar fluxes between the three ${\rm O_2}$ bands. For example, in the case of a planet orbiting a M1V star, the ratio of stellar fluxes between the B and ${\rm NIR}$ ${\rm O_2}$ bands will be much smaller than in the case of a planet orbiting a M9V star, as illustrated in Figure 3. More specifically, we used stellar models from the \cite{Husser2013} high-resolution spectral library with effective temperatures of 3600K (M1V), 3000K (M4V), and 2300K (M9V). For all the models we assumed solar abundances and ${\rm logg}$ = 4.5 dex. For the telluric spectrum we used the spectrum described in section 2.1.2 of RLM14.

\begin{figure*}[h!t]
\centering
\includegraphics[trim=1.0cm 5cm 1.5cm 2cm,clip=true,width=\textwidth]{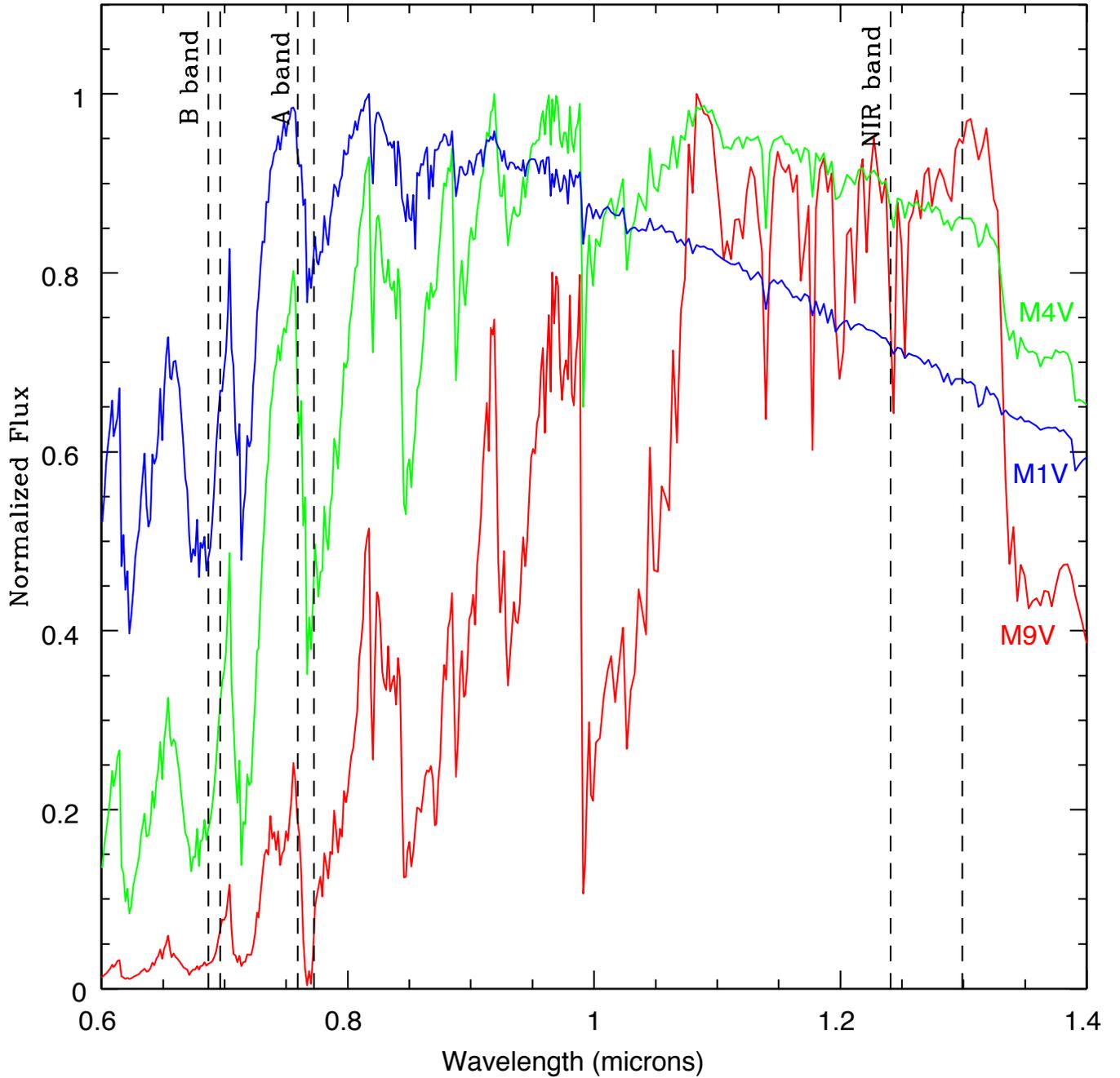}

\caption{\cite{Husser2013} model emission spectra for M1V, M4V, and M9V stars. The resolution of the models has been binned down to R=3000 for a clearer display. The vertical dashed lines highlight the location of the ${\rm O_2}$ B, A, and ${\rm NIR}$ bands. \label{fig:modelsplot}}
\end{figure*}

 The exoplanet spectrum was generated as described in section 2. For the instrumental profile ${\rm G}$ we assumed a Gaussian with a FWHM equal to the instrument's spectral resolution (in this case ${\rm R = 3\times10^5}$). We also assumed a fixed system velocity of ${\rm v_{*} = 20~km~s^{-1}}$, 
and ${\rm \epsilon}$ values of 125,000, 35,000, and 4,000, respectively, for the M1V, M4V, and M9V host star cases as listed in Table 1 of RLM14. Finally, we explored the ideal case in which only white (Poisson) photon noise is present in the observations, and the more realistic case in which the observations contain some level of red (correlated) noise in addition to white noise. Specifically, we used typically measured red noise levels of 20${\rm \%}$ in the visible \citep{Pont2006} and 50${\rm \%}$ in the near-infrared \citep{Rogers2009}.

\subsection{Simulations setup}

We ran simulations aimed at answering the following two questions: 1) Does combining observations in the ${\rm A}$, ${\rm B}$, and ${\rm NIR}$ bands significantly reduce the number of transits needed to produce a ${\rm O_2}$ detection? and 2) Does the optimal band to detect ${\rm O_2}$ depend on the spectral type of the host star? The answers to those questions are illustrated in Figures 4 and 5. 

To generate Figures 4 and 5, we estimated the photon flux per transit assuming transit durations of 4.0 h, 2.1 h, and 0.43 h for an Earth-like planet in the middle of the habitable zone of a M1V, M4V, and M9V star. To compare our results with those in RLM14, we assumed the observations are performed with an instrument similar to G-CLEF on the GMT \citep{Szentgyorgyi2012}, and assumed duty cycles of 0.93 for the M1V star, and 0.98 for the M4V and M9V stars. These duty cycles were computed using the most recent exposure time calculator (ETC) for G-CLEF. In the new ETC, observations of the M1V star  system have a higher duty cycle than the 0.86 value used in RLM14. The duty cycles for the M4V and M9V stars remained the same. The calculations also assume the same amount of flux is collected from the star at $R = 3 \times 10^{5}$ and $R = 1 \times 10^{5}$ \footnote{In other words, we assume no additional slit losses following the logic that this can be achieved by using pre-slit optics as discussed in RLM14. This assumption allows us to compare how the efficiency of ${\rm O_2}$ observations would improve with resolution.}, and  use a readout (dead) time of 7 s and a readout noise of 5 electrons per exposure, based on the most up to date values for those parameters. Like in RLM14, we assumed a dead
time of 10 s and a read-out noise of 10 electrons for a hypothetical detector in the NIR. We also set a maximum exposure time of 600 s (200 s in the case of the M9V host star), so the radial velocity shift of the remote planet does not smear out the absorption lines in its atmospheric spectrum.

Finally, to estimate uncertainties in the achieved detection levels (or equivalently, in the number of transits needed), we generated 1000 model spectra with randomized noise values and used those as a pool of transit spectra for a bootstrap analysis. In that bootstrap analysis we simulated 
transits by randomly selecting N spectra from the 1000 models pool, where N is the number of in-transit observations, and produced an observed planet spectrum as ${\rm P_{obs} = C_{in} / C_{out}}$, where ${\rm C_{in}}$ is the integrated in-transit spectra and ${\rm C_{out}}$ is the integrated spectra observed out of transit. ${\rm C_{out}}$ is produced by integrating N random spectra of the star and tellurics alone, also with randomized noise values. ${\rm P_{obs}}$ is then cross-correlated with a noiseless, template exoplanet transmission spectrum. We determine the detection significance level ${\rm \sigma}$ by examining the significance of the peak in the cross-correlation function at  ${\rm v_* = 20~km~s^{-1}}$, which is the input velocity of the system in the models. We repeated this process 500 times for each set of {\it number of transits} in Figures 4 and 5. The errorbars of each data point in those figures correspond to the 1${\rm \sigma}$ standard deviations of those 500 iterations.
\begin{figure*}[h!t]
\centering
\includegraphics[trim=1.0cm 5.7cm 1.7cm 3cm,clip=true,width=\textwidth, height=17cm]{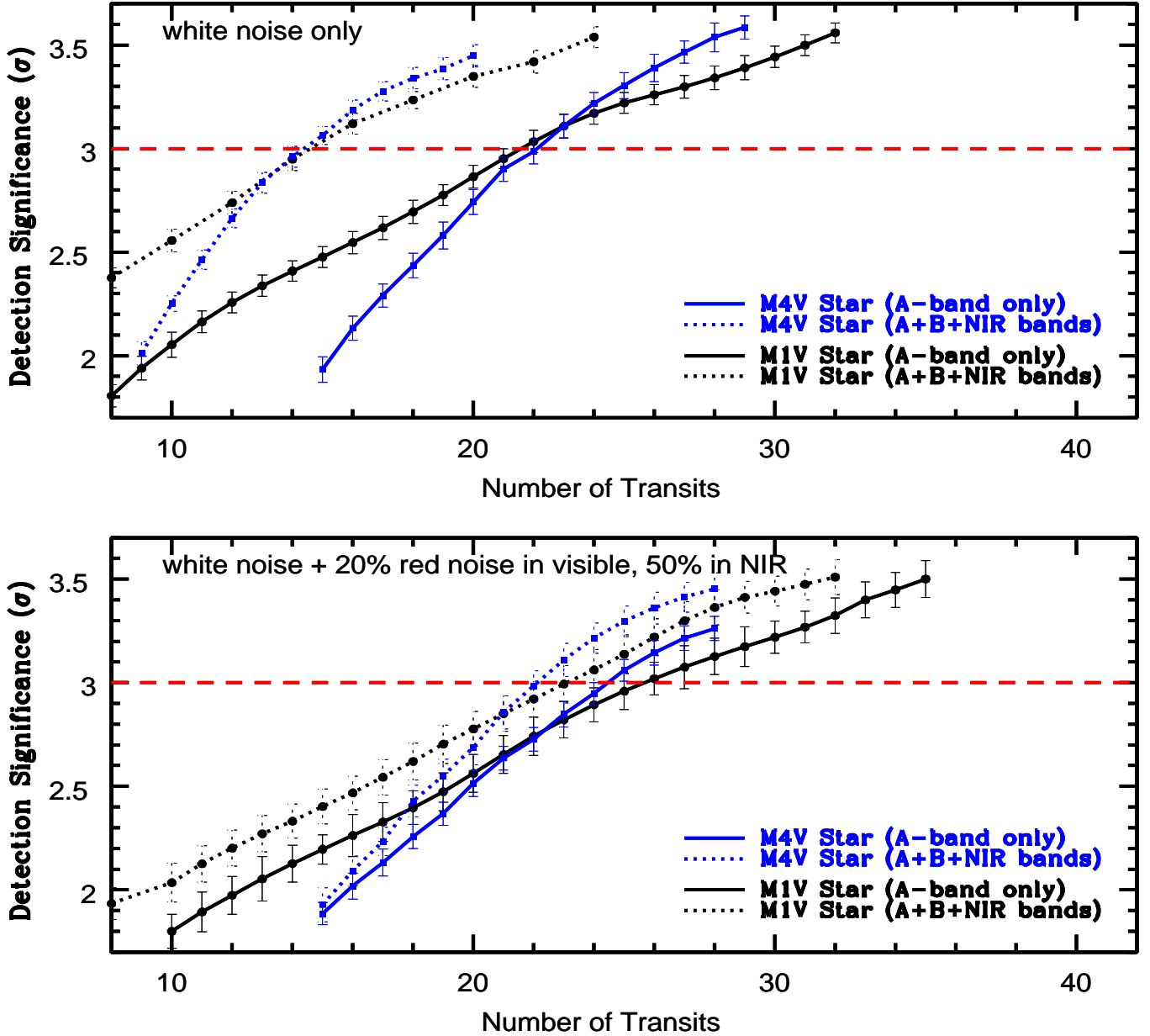}
\caption{
Number of transits needed to reach a 3${\rm \sigma}$ detection using only the ${\rm O_2}$ A-band (solid lines) versus combining the A, B, and NIR bands (dashed lines). Blue is for a M4V host star. Black is for a M1V host star. The top panel only includes white noise. The bottom panel includes 20$\%$ red noise in the optical bands, and 50$\%$ in the NIR band. \label{fig:fig4}}
\end{figure*}

\subsection{Simulations results}

Figure 4 illustrates the number of transits necessary to produce a 3${\rm \sigma}$ detection of oxygen for a planet around a M1V and a M4V star, when only using the ${\rm O_2}$ A-band versus using the ${\rm A}$, ${\rm B}$, and ${\rm NIR}$ bands combined. The case of a planet around a M9V star requires over 200 transits to produce a 3${\rm \sigma}$ detection and therefore has not been included in the plot. In the white noise only case, we find that we need 22 $\pm$ 1 transits to produce a 3${\rm \sigma}$ detection of oxygen when only using the A-band. We obtain the same number of transits for both stars, consistently with the same simulation in RLM14. However, the total number of transits needed is 34$\%$ less than in RLM14 because of the higher spectral resolution of our simulations. We find that combining the ${\rm A}$, ${\rm B}$, and ${\rm NIR}$ bands is advantageous in the case where the observations are only affected by white noise, with the required number of transits dropping to only 14 $\pm$ 1 when combining the three bands. That represents a 36 $\pm$ 5 $\%$ improvement in the number of transits compared to the results using only the A-band. When we include typically observed levels of red noise in the simulations, we find that the number of transits needed to produce a 3${\rm \sigma}$ detection of oxygen using only the A-band is 24-26 transits, slightly larger than in the white noise only case, which is expected. However, in this case the number of transits needed to produce a 3${\rm \sigma}$ detection of oxygen does not significantly decrease when combining observations in the three ${\rm O_2}$ bands. In fact, doing so only reduces the number of transits by 10 $\pm$ 8 $\%$. Therefore, there is no significant advantage in combining observations in the A, B, and NIR bands versus observing only in the A band when typical levels of red noise are present. It is worth mentioning that we also find that the contribution of the B-band is very small (as can be also inferred from Figure 5), and including it in combination with the A and NIR bands only improves the number of transits by 0.5 transits versus the results from only combining the A and NIR bands. We show the three bands in Fig. 4 for completeness.

Figure 5 shows the number of transits necessary to produce a 3${\rm \sigma}$ detection using individual ${\rm O_2}$ bands for planets orbiting M1V, M4V, and M9V host stars. The simulation results for the
M9V host star case have not been included in the A-band and B-band panels because they yield impracticably large numbers of transits. The reason for this is the very low amounts of flux M9V stars emit at those wavelengths, as illustrated in Figure~\ref{fig:modelsplot}.

The main conclusion from Fig.~5 is that, in the most realistic case of datasets with typical levels of red-noise, observations in the A-band will be the most efficient, for all low-mass host stars, except for M9V stars, since for those stars the flux emitted in the A-band is practically zero (See Fig. 3). For planets orbiting around M9V stars, ${\rm O_2}$ searches are only feasible using the NIR band, but will require scores of transits, as illustrated in the bottom panel of Fig.~5. The middle panel in the figure shows how it will be necessary to observe several hundreds of transits to produce a 3$\sigma$ detection of ${\rm O_2}$ in the B band. This is both because the B band is shallower than the A and NIR bands, and also because M dwarfs emit less flux in that band than at the wavelengths where the A and  NIR bands are located (see Fig.~\ref{fig:modelsplot}). Therefore, the B band is not well suited for observations of ${\rm O_2}$ in planets around M dwarf stars.

\begin{figure*}[h!t]
\centering
\includegraphics[trim=1.0cm 5.7cm 1.7cm 3cm,clip=true,width=\textwidth, height=20cm]{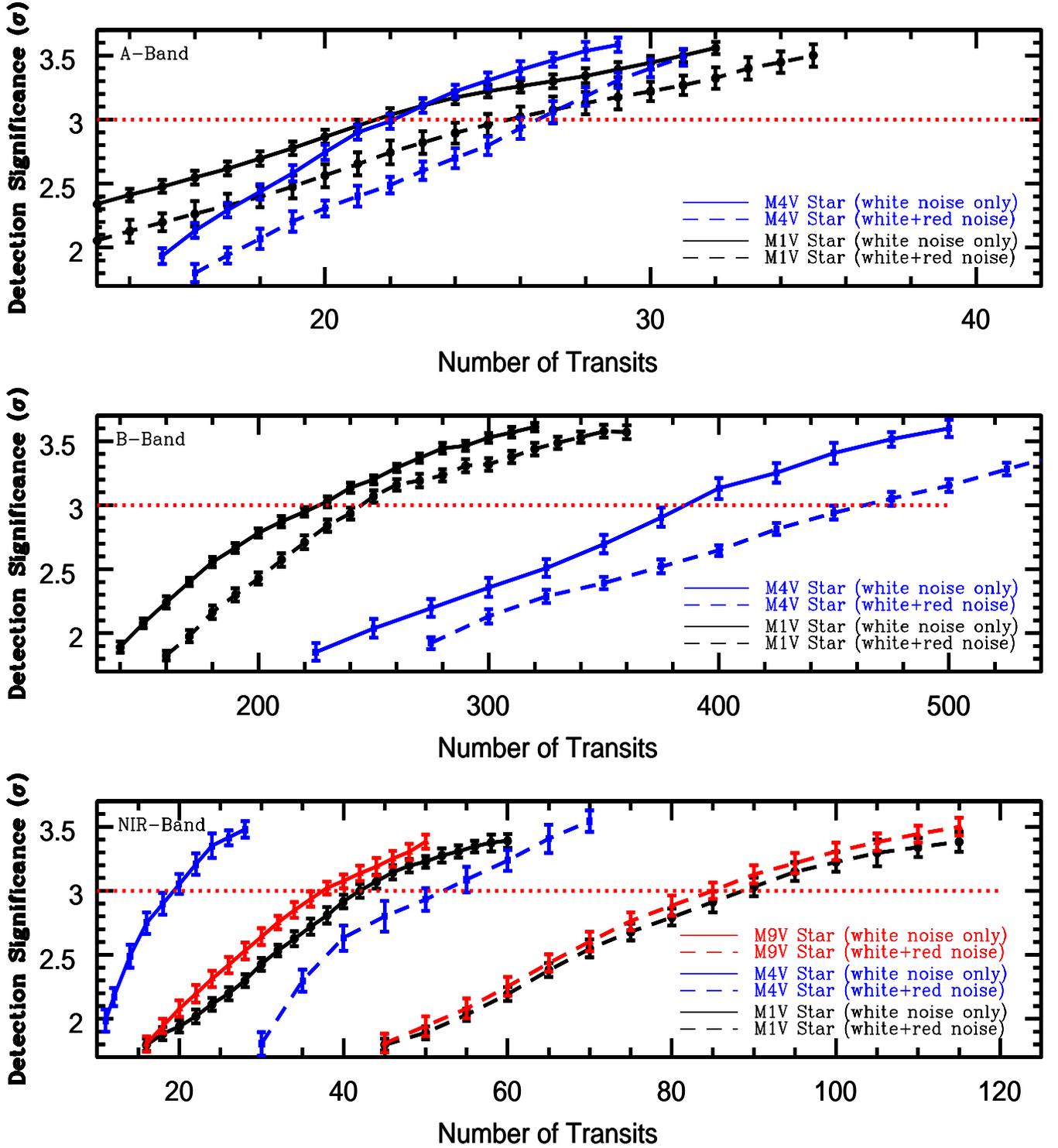}
\caption{
Number of transits needed to reach a 3${\rm \sigma}$ detection using only the ${\rm O_2}$ A-band (top panel), B-band (middle panel) and NIR-band (bottom panel), assuming host stars of spectral types M1V, M4V, and M9V. The solid lines show results when we only include white noise in the simulations. Dashed lines show results when we include red noise. The simulations using a M9V host star produce very large numbers of required transits in the case of the A and B bands, and are therefore not included in those panels.}
\label{fig:fig5}
\end{figure*}

Finally, we elaborate on the results in the bottom panel of Fig.~\ref{fig:fig5}, for the NIR band. Because our simulations assume a fixed distance of 5 pc for all the systems, the amount of flux received from each system correlates with the spectral type of the host star. This, combined with the emission spectrum of each star (Fig.~\ref{fig:modelsplot}), yields some interesting results. We find that, for the case of an M4V host star, observations in the NIR could be slightly more efficient than observations in the A band if it were possible to obtain red-noise free observations (see solid blue lines in the top panel of Fig.~\ref{fig:fig4} and the bottom panel of Fig.~\ref{fig:fig5}). We also find that, if two systems hosted by a M1V and a M9V star were found at the same distance from Earth, both would require about the same number of transits in the NIR band to produce a ${\rm O_2}$ detection. This is because the extra flux of the M9V host star at those wavelengths would get compensated by the higher intrinsic brightness of the M1V host star. However, it would still be about 3.3 times more efficient for the system around a M1V host star to observe it with an instrument in the visible covering the ${\rm O_2}$ A band, as illustrated in the top panel of Fig.~\ref{fig:fig5}.

\section{Relative Velocity and Sky Coordinates} \label{sec:velocity}

High resolution, Doppler shift searches for ${\rm O_2}$ (or any other molecule) in Earth-like planets take advantage of the relative velocity of the exoplanet system with respect to Earth, ${\rm v_*}$. Given high enough resolution, the ${\rm O_2}$ lines from the exo-Earth can be separated from the telluric ${\rm O_2}$ lines if they appear sufficiently Doppler shifted. One obvious conclusion from this method is that it will not work for exoplanetary systems with ${\rm v_* = 0}$, because the exoplanet's ${\rm O_2}$ lines will fully overlap with telluric ${\rm O_2}$ lines. However, as first pointed out by RLM14, this line blending problem is not unique to ${\rm v_{*} = 0~km~s^{-1}}$ and there is other relative velocities at which blending effects will be significant, given the high density of ${\rm O_2}$ lines.

In this section we refine the blending analysis by RLM14 for the ${\rm O_2}$ A band and NIR band by using a finer grid of relative velocities (${\rm \Delta v_* = 0.5~km~s^{-1}}$),  and also considering a larger number of ${\rm O_2}$ lines. In addition, we update the blending criterium used in RLM14 by counting as ${\it blended}$ all the exoplanet lines with a separation from a telluric line of less than one resolution element. The reason for excluding those lines is that, even though some of them would still appear in the exoplanetary spectrum after substracting the telluric lines, the signal-to-noise level in those portions of the exoplanet spectrum would be too low to detect those lines. Finally, we extend the line blending analysis here to resolutions of ${\rm 3\times10^5}$ and ${\rm 5\times10^5}$, following the conclusions in sections 2 and 3 that higher resolutions increase the depth of the ${\rm O_2}$ lines and therefore their detectability. 

Figure 6 shows the normalized fraction of blended lines as a function of the relative velocity of the exoplanetary system with respect to Earth, for the ${\rm O_2}$ A band and NIR band. The figure, which shows blending fractions for resolutions ${\rm 1\times10^5}$, ${\rm   
3\times10^5}$, and ${\rm 5\times10^5}$, illustrates how there will always be a non-zero level of blending between the exoplanet lines and the telluric lines. Blending is higher in the A-band than in the NIR band, because of the higher density of lines. However, as shown in sections 2 and 3, the A-band is still the best suited for ${\rm O_2}$ observations for the majority of systems, because of deeper absorption lines, and higher stellar emission levels for all host stars except M9Vs. 

Focusing on the A-band, our results overall agree with RLM14. Systems with relative velocities between about ${\rm \pm~15 - 30~km~s^{-1}}$ are the least affected by line blending. The figure also shows how increasing the spectral resolution reduces the level of blending. In addition, systems with negative relative velocities with respect to Earth will be on average less affected by blending than systems with positive relative velocities. This is because of the asymmetric shape of the ${\rm O_2}$ A-band, and also of the individual lines. Also this effect increases with resolution. To quantify this effect we calculated the average blending between -150 ${\rm km~s^{-1}}$ and -15 ${\rm km~s^{-1}}$, and between +15 ${\rm km~s^{-1}}$ and +150 ${\rm km~s^{-1}}$ for the three blending functions shown in the top panel of Figure 6. The results show that, for R= ${\rm 5\times10^5}$, systems with negative relative velocities have on average 5.95$\%$ less blending of lines than systems with positive relative velocities. In the case of R= ${\rm 3\times10^5}$ the blending difference is 5.46$\%$, while for R= ${\rm 1\times10^5}$ the blending difference is 1.95$\%$. In the case of the NIR band, which has a more symmetric shape, this effect is not noticeable.


Although blending is unavoidable in the search for ${\rm O_2}$ in an exoplanetary atmosphere when observing from the ground, it can be reduced by combining information about the systemic velocity of the exoplanetary system, with their sky coordinates, and with the position of Earth in its orbit during the observations. Here we produce a quantitative estimation of which exoplanetary systems will be best suited to search for ${\rm O_2}$ in their atmospheres based on their  sky coordinates and changes in ${\rm v_*}$. As shown in the top panel of Figure~\ref{fig:Fig7}, the relative velocity of a system with respect to Earth can be written as
\begin{equation}
{\rm v_{*}} = {\rm v_{sun,*}} + {\rm v_{Earth}} \cdot {\rm sin}{\theta}\cdot {\rm cos}{\phi} ,
\end{equation}

\noindent where ${\rm v_{sun,*}}$ is the relative velocity between the exoplanet system and the Sun, ${\rm v_{Earth}}$ is the velocity of Earth around the Sun, ${\rm \phi}$ is the angle between the position of the exoplanet host star in the sky and the ecliptic, and ${\rm \theta}$ is the angle between the location of Earth on the ecliptic at a given epoch, and the projected line of sight between the location in the sky of the exoplanet host star and the Sun.

\begin{figure*}[h!t]
\centering
\includegraphics[trim=1.0cm 5.5cm 2.5cm 3cm,clip=true,width=\textwidth]{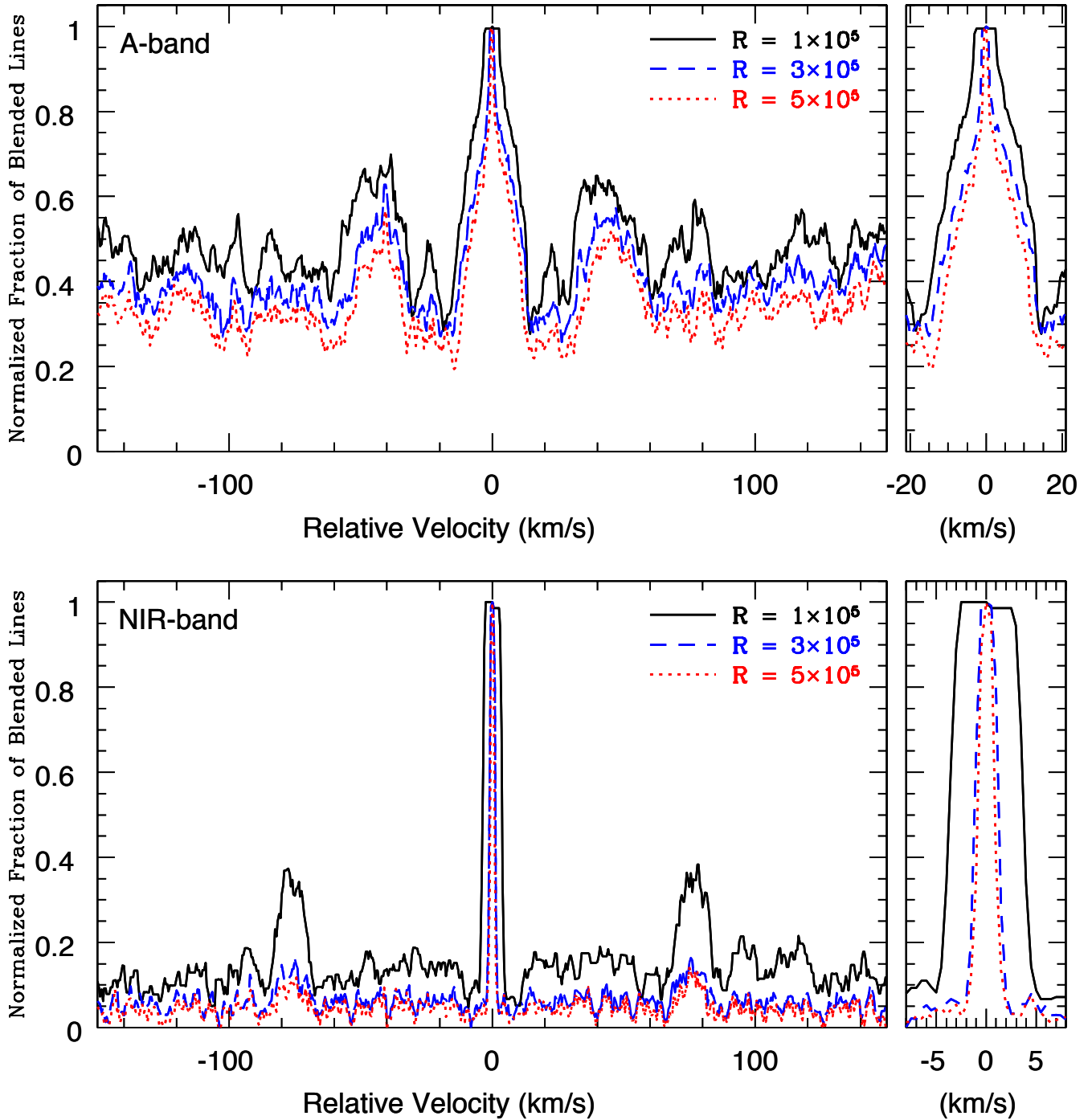}
\caption{{\it Top} -- Normalized fraction of blended lines in the ${\rm O_2}$ A band as a function of the relative velocity of the exoplanet system with respect to Earth, for three spectral resolution values (${\rm R = 1\times10^5}$, ${\rm   
3\times10^5}$, and ${\rm 5\times10^5}$). The right-side panel shows a zoom-in around  ${\rm v_{*} = 0~km~s^{-1}}$. {\it Bottom} -- Same for the ${\rm O_2}$ NIR band.}\label{fig:Fig6}
\end{figure*}

\begin{figure*}[h!t]
\begin{center}$
\begin{array}{c}
\includegraphics[trim=1.0cm 2.0cm 2.5cm 3.0cm,clip=true,width=\textwidth]{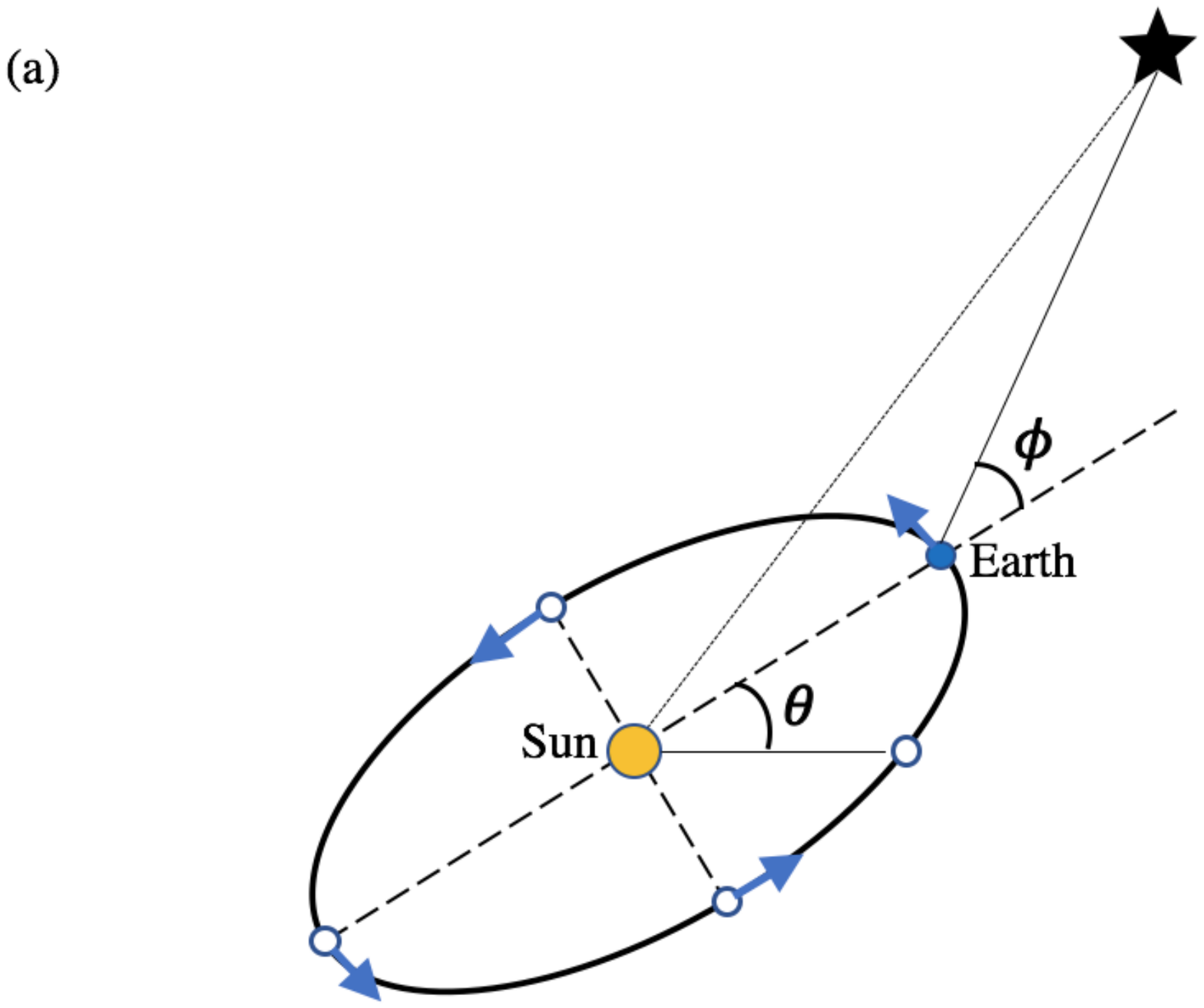}\\
\includegraphics[trim=1.0cm 15.35cm 1cm 3.5cm,clip=true,width=\textwidth]{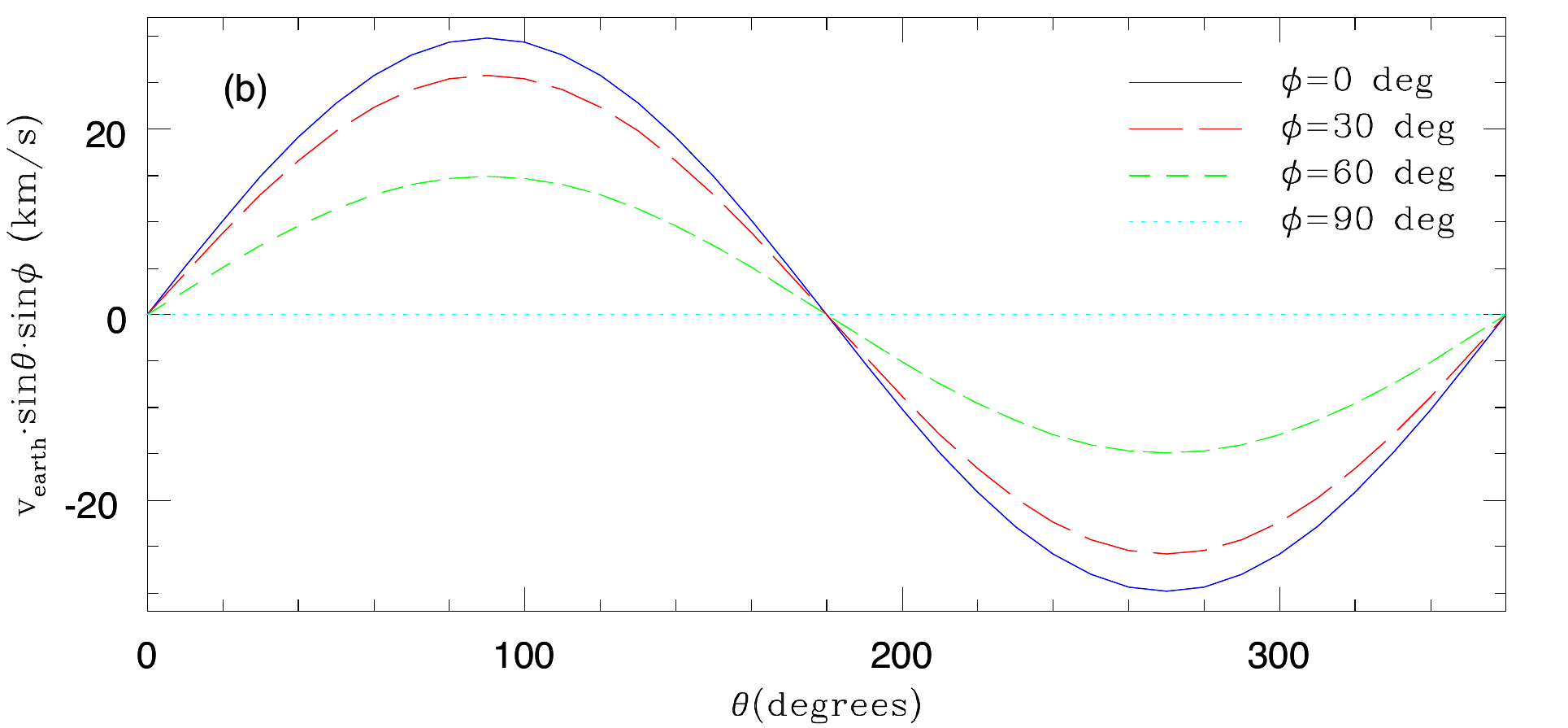}
\end{array}$
\caption{${\it Top}$ -- Schematic diagram to illustrate the derivation of eq.~5 in the text (see description of the parameters shown in the figure in section 4). ${\it Bottom}$ -- Variation in the projected velocity of Earth during its orbit around the Sun for different position angles $\phi$ with respect to the ecliptic.}
\label{fig:Fig7}
\end{center}
\end{figure*}

The bottom panel of Figure~\ref{fig:Fig7} shows how the relative velocity of the exoplanetary system changes relative to Earth, as Earth moves in its orbit around the Sun, for different values of $\phi$. If the exoplanet host star is located perpendicular to the ecliptic (${\rm \phi}$ = 90 deg), its velocity relative to Earth will remain constant. In this case, systems with ${\rm v_*}$ values that coincide with high blending peaks in Fig.~\ref{fig:Fig6} are not optimal, and priority should be given to systems with ${\rm v_*}$ values with low blending levels. If  the exoplanet host star is located along the ecliptic (${\rm \phi}$ = 0 deg), its velocity relative to
Earth will very by 59.6 ${\rm km~s^{-1}}$ periodically through the year, as Earth moves along its orbit. In this case, the blending levels of a given system will change periodically, so observers can optimize observations by targeting the system during epochs when blending is reduced. The same is true for systems with intermediate $\phi$ values, for which the amplitude of their ${\rm v_*}$ variation will depend on their actual sky coordinates.

\section{Summary and Discussion} \label{sec:discussion}

We have generated detailed simulations of observations of ${\rm O_2}$ in the atmosphere of a hypothetical nearby transiting Earth-twin using large ground-based telescopes. Our goal is to establish informed guidelines for the design of instrumentation and observational approaches best suited to search for ${\rm O_2}$ in the atmospheres of nearby potentially Earth-like planets. We find that performing the observations at resolutions of  ${\rm R = 3-4 \times10^5}$, instead of the standard  ${\rm R \sim 1 \times10^5}$ of most current and planned high-resolution spectrograph designs, more than doubles the depth of the exoplanet's ${\rm O_2}$ lines. That increase in line depth translates into a reduction of over 30$\%$ in the number of transits needed to produce a detection.  For our calculations we have assumed the atmosphere of the simulated exoplanet has a Pressure-Temperature (P--T) profile and an ${\rm O_2}$ atmospheric vertical distribution identical to Earth. In reality this might not be the case: we might find planets with different P--T profiles or  with different atmospheric vertical distributions of ${\rm O_2}$. In those cases the ${\rm O_2}$ lines will appear intrinsically broader or still not fully resolved at ${\rm R = 3-4 \times10^5}$, but given the current absence of other examples, using Earth as a proxy seems justified. Also, observing at higher resolutions than the currently planned ${\rm 1 \times10^5}$, might yield additional information about ${\rm O_2}$ in exoplanetary atmospheres by enabling more detailed sampling of the absorption lines.

We also find that given the relative depths of the four ${\rm O_2}$ absorption bands in the optical to near-infrared wavelength range, the relative emission of M0V to M9V host stars in those bands, and typical levels of correlated noise for current ground-based instrumentation/observations, there is no advantage to performing combined observations in multiple ${\rm O_2}$ bands and the A-band remains the best suited band to search for ${\rm O_2}$. Therefore, based on our current knowledge, it appears that future efforts to detect ${\rm O_2}$ in the atmospheres of transiting exoplanets are best focused on optimizing instrumentation and observational approaches for the A band. Designing new instrumentation which optimizes ${\rm O_2}$ A-band observations \citep[see][]{BenAmi2018}, and maybe also building dedicated spectrographs with optimize throughput in the ${\rm O_2}$ bands will be key. The recommendation to use only the ${\rm O_2}$ A-band, however, should be revised if/when we manage to further reduce red-noise levels in the observations (especially in the near-infrared), since our simulations show how combining observations in different bands will be advantageous if red-noise levels drop. One caveat of our simulations is that they use red noise levels found in optical and near-IR time series photometry. However, it is not known whether high resolution, time series observations at wavelengths covering ${\rm O_2}$ bands present similar levels of red noise. Such observations would be useful. In addition, improvements in detector noise will be have a great impact in this particular science case, and also others.

Because of the high density of lines in the ${\rm O_2}$ bands, ground-based observations will always suffer from some level of blending with telluric ${\rm O_2}$ lines. The problem is most pronounced for systems with relative velocities less than about $\pm$ 15 ${\rm km~s^{-1}}$ and between $\pm$ 35-55 ${\rm km~s^{-1}}$ with respect to Earth. We find that, as expected, increasing the spectral resolution of the observations decreases the amount of blending, and therefore increases the number of systems that can be observed. We also find that it will be possible to observe some systems with high blending, if their sky coordinates of and epoch of the year are taken into account when planning the observations. 
It is important to take these limitations introduced by blending with telluric lines into account, especially in the case of NASA TESS mission candidates. The continuing viewing zone (CVZ) of TESS is centered around the ecliptic poles. Therefore we expect to find most of the best primary mission, planet candidates in those locations as illustrated for example in Figure 4 of \cite{Huang2018}. Also, regions within $\pm$ 6 deg from the ecliptic plane will not be observed during TESS's primary mission. Furthermore, the CVZ of JWST also points towards the ecliptic pole, so the same {\it best} targets from TESS will be the focus of follow-up observations in search for potential biomarkers, such as ${\rm H_2O}$ or ${\rm CH_4}$ from space. We suggest that the observability of ${\rm O_2}$ should be included as a key criterium when ranking those candidates for atmospheric follow-up. Blending with telluric lines can be eliminated by observing from space, but the need for very large photon collecting areas currently limits ${\rm O_2}$ searches to ground-based ELTs. 

Another factor that needs to be determined is the stability of telluric lines, in particular how stable are the wings of telluric lines at the resolution and time scales required for these observations. Like in the case of the red noise levels for high resolution observations mentioned above, observations to establish how stable telluric lines are will be useful in the design of future observations of ${\rm O_2}$.

Finally, this and previous studies show how the expected exoplanet ${\rm O_2}$ signals are very small. Even for systems just a few parsecs away from Earth, and with the large photon collecting areas of ELTs, we will still need to co-add observations from many transit events of an individual system to produce a significant  ${\rm O_2}$ detection. This problem could be improved by observing, whenever possible, the same targets with more than one ELT. Coordinating and combining observations of the same targets with different ELTs will not only lead to a faster collection of the necessary amount of photons to produce a detection, but it will also serve as cross-check for spurious signals or systematics.




\acknowledgments
The authors thank the Brinson Foundation and the Smithsonian Institution for providing funding to support this project. We also thank the anonymous referee for helpful comments  and suggestions. JGM  acknowledges  support  by  the  National  Science Foundation through a Graduate Research Fellowship under grant No. DGE1745303 and by the Ford Foundation through a Ford Foundation Predoctoral Fellowship, administered by the National Academies of Sciences, Engineering, and Medicine.

\end{document}